# Study on the GEANT4 code applications to the dose calculation using imaging data

**Jeong Ok Lee and Jeong Ku Kang**

*Wonkwang Health Science University, Iksan 570-750*

**Jhin Kee Kim, Hyeong Cheol Kwon, Jung Soo Kim and Bu Gil Kim**

*Chonbuk National University Hospital, Jeonju 561-712*

**Dong Hyeok Jeong**

*Dongnam Inst. of Radiological & Medical Sciences, Busan 691-953*

The use of GEANT4 code has increased in the medical field. There are various studies to calculate the patient dose distributions with the GEANT4 code using the imaging data. In present study, the Monte Carlo simulations based on the DICOM data were performed to calculate absorbed dose in the patient's body. Various visualization tools were equipped in the GEANT4 code to display the detector construction, however there are limitations to display the DICOM images. In addition, it is difficult to display the dose distributions on the imaging data of the patient. Recently, gMocren code, volume visualization tool for GEANT4 simulation, has been developed and used in volume visualization of image files. In this study, the imaging data based absorbed dose distributions in patient were performed by using the gMocren code. The dosimetric evaluations with TLD and film dosimetry methods were carried out to verify the calculation results.

Email: kimjk@jbnu.ac.kr

Fax: +82-63-250-1192

# I. INTRODUCTION

GEANT4 is a toolkit used to simulate the passage of particles through matter. Recently, it has been used in many medical physics applications. As one of the cancer treatment, the radiotherapy has been adopted. Since a surgical operation is not needed, a burden of human body is small in the radiotherapy. Moreover, a shape and function of the internal organ can be also kept. However, the determination of a suitable irradiation range is difficult because we don't treat a cancer directly. In radiotherapy, GEANT4 has been applied to accurately simulate the propagation of particles and the interaction of particles, not only with medical devices, but also with patient's phantoms[1,4]. The use of GEANT4 code has increased in the medical field. And many researchers try to use patient's image data to calculate the dose. It's considered to be more accurate method to calculate dose in radiotherapy than simple codes. We construct detector with parameterized volume for GEANT4 simulations, which can be applied to simulations using medical image data as the input. We tried to apply this code to the patient's medical images to simulate the propagation and interaction of the particles. So we can calculate the absorbed dose of the patient. In this study, the used visualization tool is called gMocren[1,2]. We used dicomHandler class object to convert from the dicom binary file format to the ASCII file format which contains the matrix number, pixel size and pixel's HU. One of the contributions of this study was the identification of the pathway toward the gMocren, as well as the investigation of the related image structures and internal processes, which enabled a better understanding of medical image and resulted in developing intervention methods to counteract isodose distribution of image data. The purpose of this present paper is to extract the dose table in an analyzed data of dose distribution, using real patient's medical image data with a program based on Monte Carlo simulation and visualization tool for radiation isodose mapping.

## II. EXPERIMENTS AND DISCUSSION

DICOM (Digital Imaging and COmmunication in Medicine) interface handler that provided from GEANT4 doesn't work correctly, so we had been modified independently for PHILIPS, SIMENSE CTSim medical image data and had been tested. Intermediate file creation modules are plugged in for converting original CT data into a suitable one for making the geometry. It's extraction of patient geometry, reformation of voxels, and calculation of density from DICOM file's pixel data. Each file corresponds to a Z slice. The Z slices will be merged at runtime to form a unique patient volume. The DICOM images pixel values represent CT-Hounsfield numbers and they should be converted to a given density and then to a material type. The relation between CT number and density is more or less linear. We use machine specific calibration curve to convert CT-Hounsfield number to physical density for attenuation effect in beam interaction. An assignment of material densities to materials are done from the information in the file DATA.dat. It is a matching progress to convert image data for using DICOM tool in GEANT4 simulation. The data taken from the International Commission on Radiation Units and measurements (ICRU) report 46 was used to build the materials (lung, liver, breast, bones). We have converted DICOM header information data of the PHILIPS CT image data into a suitable format for making the geometry.

gMocren is very useful analysis tool for GEANT4 simulation with DICOM files. gMocren shows a screen, which consists primarily of the following panes; a volume rendering pane, a multiplanar reformat (MPR) pane, and two histogram panes. The volume-rendering pane displays three-dimensional (3D) images. The MPR plane displays three two-dimensional (2D) images, which are the projections of the 3D image onto the xy, yz, and zx planes, respectively. In addition, two histogram panes allow the user to adjust transfer functions of patient data and a dose distribution. In addition to these panes, several icons can be used for saving an image or changing image contrast, resolution and brightness. The existing driver of the GEANT4 toolkit can visualize the simulated trajectory of materials and particles. Also, gMocren, with its volume rendering technique, can display both the patient data and the simulated dose distribution. We have calculated the absorbed dose of patient with real patient's image files and displayed this dose with gMocren code.

We calculate the absorbed dose of patient with real patient's image files and display this dose with gMocren code. We planned same patient with Radiation Treatment Planning System (RTPS) which is used in radiation therapy planning (XiO). We attempted to compare simulation data with real planned data. But it is difficult to display the gMocren data, dose.gdd, like a planned dose data. gMocren has the utility files like dump.gdd of routine process. The dose distribution and image information in GEANT4 had previously been stored in root format (dicom.out) and gMocren format (dose.gdd). However, we can't create a file that we use to compare with every point. gMocren must be revised for quantitative analysis routine to use more commonly.

In results and discussion, we calculate the absorbed dose of each voxel point in every slice for real patient's image files and compared with measured dose and RTP's planned dose[9,10]. Our result displayed a calculated dose to the computed tomography and reviewed with gMocren tools, which is shown in Figure 1. and Figure 2. And compared with RTP's plan data in field size 10x10 cm$^2$ are shown in Figure 3. As can see from Figure 1, the user can intuitively and easily obtain and verify the simulated dose distribution with this visualization tool. In order to reduce the time required for rendering a 3D image, the ray-casting method and the early ray termination method were adopted, resulting in the accurate visualization of both patient data and dose distribution. In addition, the user can rotate, zoom in or out, and move the object. By using the two histograms, the user can change opacities and colors corresponding to a given value on patient or dose distribution panes. For example, if the user sets a suitable opacity curve in the patient histogram of gMocren, then the computed tomography data and the dose distribution for the pelvic are shown in Figure 2. And the simulated same CT data and compared RTP's plan data in field size 10x10 cm2 shown in Figure 3. These images are displayed with the (A) xy, (B) yz, and (C) zx plane views and compared with the Figure 1. It's inconvenient that it can't compare each point doses and display isodose line like RTPS plan data. Nevertheless, it's very useful program and can use freely.

GEANT4 is extensively applied to radiology because of its capability to handle all particles including ions, complex geometry and electromagnetic fields and flexibility. The software suit for simulating radiotherapy has been developed in the several fields as the application of software for simulation in radiotherapy. The 2D image is displayed using three sectional views with MPR. Pointing with the mouse displays the dose value, while the cross line changes the position of the view in the two other panes changes. Using gMocren, the user can assign a proper color in gMocren, the simulated behavior of the irradiated particles can display clearly. gMocren is very useful tool for displaying the GEANT4 simulation result, but difficult to find every voxel real dose value, so we developed to create the each slice's dose matrix and stored with table for further use. The dose distribution table of the center slice was calculated. The center slice data are shown in Table1. The first columns are voxel's copy numbers and second are doses in that voxels. In order to display the simulated dose distribution, the simulated results must be stored in an appropriate file format [1, 4].

In our case, the data file consists of the following entries: the simulated dose distributions, the modality image. gMocren has a function that can be used to take a screenshot of gMocren window, which can then be saved in either a JPEG or TIFF format. The simulated dose distribution can be displayed as a shadow contour plot. But we can't create nice contour plot on the medical image like planned data with RTP system. And it is very difficult to compare with RTP's plan data. We need more study to this subject for quantities compare.

### III. CONCLUSION

In conclusion, we have applied a volume visualization tool for GEANT4 simulation. The graphic displays of the GEANT4 toolkit with gMocren are suitable for displaying either the complex patient data or the simulated dose distribution. gMocren supports a volume rendering technique and can display both complex patient data as well as dose distribution simulated using a GEANT4 simulation. But we need more convenient methods for quantitative analysis of our results. So we have developed to create the each voxel's dose tables of the every slice and analyze the distribution with DICOM file. gMocren is very convenience tool but provide only qualitative analysis. We need more enhanced functions to display contour map like RTP or utility program to create dose table in every points. Therefore, we need more study in this subject. It is considered that the analyzed Monte Carlo based simulation program might be useful to study for isodose mapping and dose distribution coverage in 3DCT volume due to variations of image information.

## ACKNOWLEDGEMENT

This paper (exhibition practice etc) was supported by Wonkwang Health Science University in 2013.

Table 1. Calculated Dose Distribution Tables of the Center Slice Data. First Columns are Voxel's Copy Numbers and Second are Doses in that Voxels.

| Voxel copy number | 442368 | 442496 | 442624 | 442752 | 442880 | ... |
|---|---|---|---|---|---|---|
| Dose (Arbitrary Unit) | $4.44 \times 10^{-21}$ | $8.19 \times 10^{-21}$ | $2.20 \times 10^{-21}$ | $3.05 \times 10^{-21}$ | $6.11 \times 10^{-21}$ | ... |

Figure Captions.

Fig. 1. 2D section images of xz-xy-yz-plane with the MPR. The simulated dose is displayed by means of a distribution map.

Fig. 2. 3D images of the pelvic obtained using the dose method in gMocren.

Fig. 3. Planning data in field size 10x10 cm$^2$ to compare with calculations.

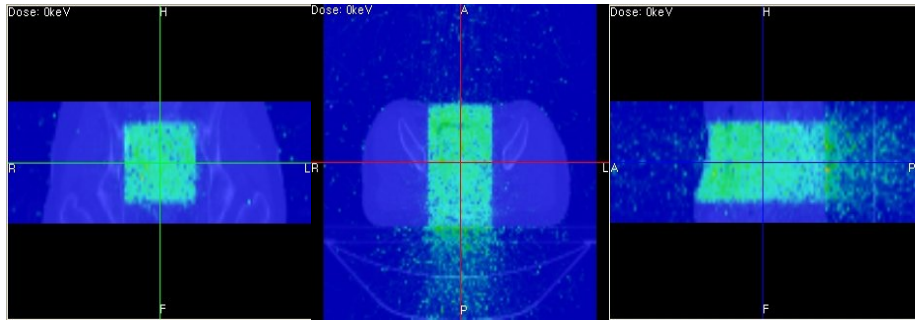

(A) *xy*　(B) *yz*　(C) *zx*

Fig. 1.

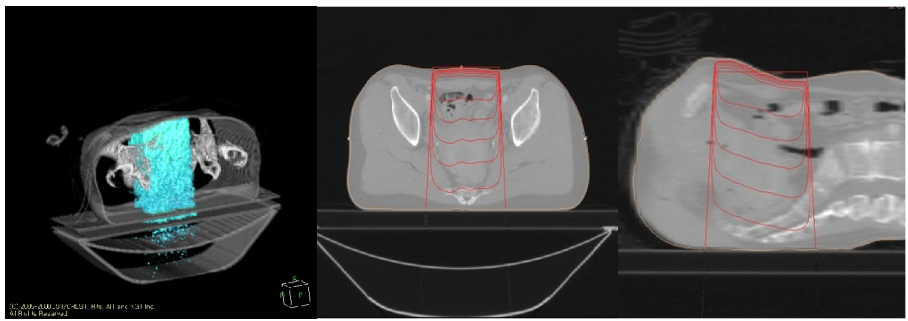

(A) *3D*　(B) *yz*　(C) *zx*

Fig. 2.

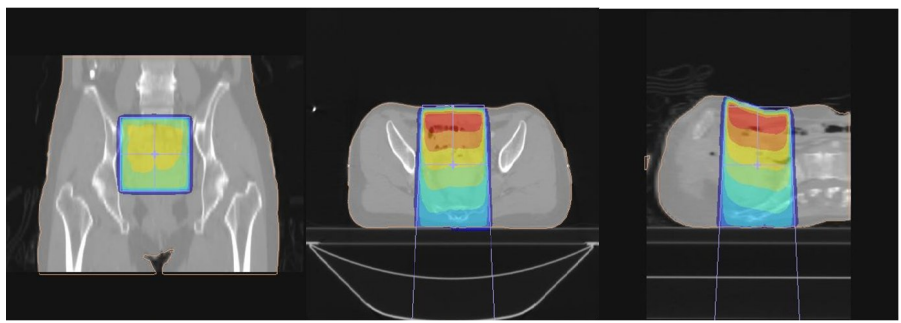

(A) *xy*　(B) *yz*　(C) *zx*

Fig. 3.